\documentclass[12pt,draftcls,onecolumn]{IEEEtran}

\makeatletter
\def\ps@headings{%
\def\@oddhead{\mbox{}\scriptsize\rightmark \hfil \thepage}%
\def\@evenhead{\scriptsize\thepage \hfil\leftmark\mbox{}}%
\def\@oddfoot{}%
\def\@evenfoot{}}
\makeatother
\usepackage{amsfonts}
\usepackage{multirow}
\usepackage[tbtags]{amsmath}
\usepackage{graphicx,subfigure}
\usepackage{color}      
\usepackage{epsfig}
\usepackage{amssymb}
\usepackage{tipa}
\usepackage{bbm}

\def\endproof{\hspace*{\fill}~$\blacksquare$}
\long\def\comment#1{}

\newcommand{\beq}{\begin{equation}}
\newcommand{\eeq}{\end{equation}}
\newcommand{\beqno}{\begin{equation*}}
\newcommand{\eeqno}{\end{equation*}}

\newcommand{\bes}{\begin{split}}
\newcommand{\ees}{\end{split}}
\newcommand{\bdm}{\begin{displaymath}}
\newcommand{\edm}{\end{displaymath}}

\newtheorem{theorem}{Theorem}
\newtheorem{claim}{Claim}

\newtheorem{definition}{Definition}

\newcommand{\bd}{\begin{definition}}
\newcommand{\ed}{\end{definition}}

\newcommand{\bv}{\begin{vugraph}}
\newcommand{\ev}{\end{vugraph}}
\newcommand{\bi}{\begin{itemize}}
\newcommand{\ei}{\end{itemize}}
\newcommand{\ben}{\begin{enumerate}}
\newcommand{\een}{\end{enumerate}}

\newcommand{\bean}{\begin{eqnarray*} }
\newcommand{\eean}{\end{eqnarray*} }
\newcommand{\bea}{\begin{eqnarray} }
\newcommand{\eea}{\end{eqnarray} }

\newcommand{\ba}{\begin{array} }
\newcommand{\ea}{\end{array} }

\newcommand{\uihat}{\hat{u}_i}
\newcommand{\bfv}{\mathbf{v}}

\newcommand{\calS}{{\cal S}} \newcommand{\calT}{{\cal T}}

\newcommand{\bfT}{\mathbf{T}}
\newcommand{\bfS}{\mathbf{S}}
 \newcommand{\bfA}{\mathbf{A}}

\newcommand{\bfP}{\mathbf{P}}

\newcommand{\di}{d_i}
\newcommand{\ui}{u_i}
\newcommand{\dihat}{\hat{d}_i}

\begin{document}

\title{Protection against link errors and failures using network
coding}
\author{\authorblockN{Shizheng Li and Aditya Ramamoorthy}
\thanks{The authors are with
Department of Electrical and Computer Engineering, Iowa State
University. Email: \{szli, adityar\}@iastate.edu}
\thanks{This material in this work appeared in part in International Symposium on Information Theory, Seoul, Korea, 2009.}
}
\maketitle
\begin{abstract}

We propose a network-coding based scheme to protect multiple bidirectional unicast connections against adversarial errors and failures in a network. The network consists of a set of bidirectional primary path connections that carry the uncoded traffic. The end nodes of the bidirectional connections are connected by a set of shared protection paths that provide the redundancy required for protection. Such protection strategies are employed in the domain of optical networks for recovery from failures. In this work we consider the problem of simultaneous protection against adversarial errors and failures.

Suppose that $n_e$ paths are corrupted by the omniscient adversary. Under our proposed protocol, the errors can be corrected at all the end nodes with $4n_e$ protection paths. More generally, if there are $n_e$ adversarial errors and $n_f$ failures, $4n_e + 2n_f$
protection paths are sufficient. The number of protection paths
only depends on the number of errors and failures being protected
against and is independent of the number of unicast connections.
\end{abstract}

\begin{keywords}
Network coding, network error correction, adversarial error,
network protection
\end{keywords}
\vspace{-1mm}
\section{Introduction}

Protection of networks against faults and errors is an important
problem. Networks are subject to various fault mechanisms such as
link failures, adversarial attacks among others and need to be
able to function in a robust manner even in the presence of these
impairments. In order to protect networks against these issues,
additional resources, e.g., spare source-terminal paths are
usually provisioned. A good survey of issues in network protection
can be found in \cite{Zhou00}. Recently, the technique of network
coding \cite{al} was applied to the problem of network protection.
The protection strategies for link-disjoint connections in
\cite{Kamal06a,Kamal07a,kamal10J} perform network coding over
p-Cycles \cite{GroverS00}, which are shared by connections to be
protected. The work in \cite{kamalR08,KRLL09} uses paths instead
of cycles to carry coded data units and proposes a simple protocol
that does not require any synchronization among network nodes, yet
protecting multiple primary path connections with shared
protection paths.
These schemes deal exclusively with link failures, e.g., due to
fiber cuts in optical networks, and assume that each node knows
the location of the failures at the time of decoding. In this work
we consider the more general problem of protection against errors.
An error in the network, refers to the alteration of the
transmitted data unit in some manner such that the nodes do not
know the location of the errors before decoding. If errors over a
link are random, classical error control codes \cite{shulin} that
protect individual links may be able to help in recovering data at
the terminals. However, such a strategy will in general not work
when we consider adversarial errors in networks. An adversary may
be limited in the number of links she can control. However for
those links, she can basically corrupt the transmission in any
arbitrary manner. An error correction code will be unable to
handle a computationally unbounded adversary who knows the
associated generator matrix and the actual codes under
transmission. This is because she can always replace the actual
transmitted codeword by another valid codeword.

In this paper we investigate the usage of network coding over
protection paths for protection against adversarial errors.
Protection against link failures in network-coded multicast
connections was discussed in \cite{rm}. The problem of network
error correction in multicast has been studied to some extent.
Bounds such as Hamming bound and Singleton Bound in classical
coding theory are generalized to network multicast in
\cite{YeungC061, YeungC062}. Several error correction coding
schemes are proposed, e.g., \cite{ZZJ08, Jaggi08, yangYNpreprint,
SilvaKK08J}. However, these error correction schemes work in the
context of network-coded {\it multicast} connections.

\begin{figure}[h]
\centerline{\psfig{figure=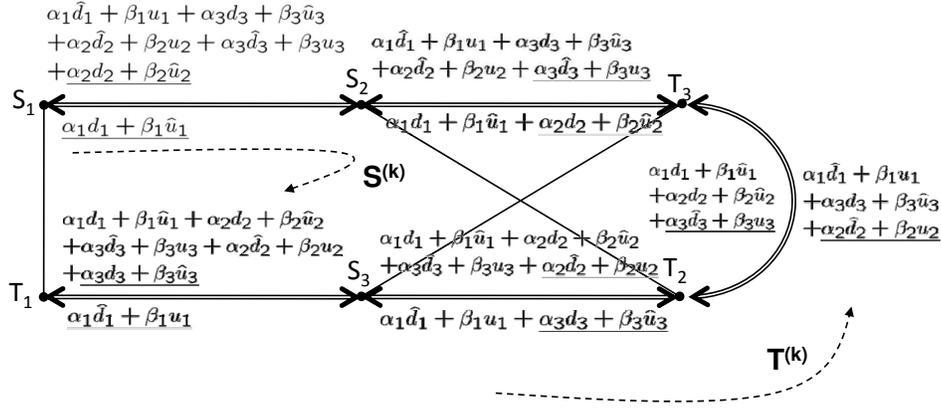,width=5in}}\vspace{-2.5mm}
\caption{Three primary paths $S_i - T_i, i = 1, \ldots, 3$ being
protected by a single protection path $\mathbf{P}^{(k)}$. The
single lines represent the primary paths and the double lines
represent the protection path. The clockwise direction of the
protection path is $\mathbf{S}^{(k)}$ and the counter clockwise
direction is $\mathbf{T}^{(k)}$. $\sigma(S_2)=T_3$,
$\tau^{-1}(T_3) = T_2$. The encoded data units on $\bfS^{(k)}$ are
labeled inside the protection path and the encoded data units on
$\bfT^{(k)}$ are labeled outside the protection path. At $T_3$,
the data unit $P^{(k)} = \alpha_1 d_1 + \beta_1 \hat{u}_1+
\alpha_2 d_2 + \beta_2 \hat{u}_2 + \alpha_1 \hat{d}_1 + \beta_1
u_1+ \alpha_3 d_3 + \beta_3 \hat{u}_3+\alpha_2 \hat{d}_2 + \beta_2
u_2$, if there is no error, $P^{(k)} = \alpha_3 d_3 + \beta_3
u_3$.} \label{fig:example-enumeration}\vspace{-2.5mm}
\end{figure}

In this work we attempt to simultaneously protect multiple unicast
connections using network coding by transmitting redundant
information over protection paths. Note that even the error-free
multiple unicast problem under network coding is not completely
understood given the current state of the art \cite{Lil04}.
Therefore we consider the multiple unicast problem under certain
restrictions on the underlying topology. In our work we consider
each individual unicast to be operating over a single primary
path. Moreover, we assume that protection paths passing through
the end nodes of each unicast connection have been provisioned
(see Figure \ref{fig:example-enumeration} for an example). The
primary and protection paths can be provisioned optimally by
integer linear programming (ILP). Although the ILP has high
(potentially exponential) computational complexity, it only needs
to run once before the transmission of data and there are powerful
ILP solvers, e.g. CPLEX, to solve ILP problems. Suppose that the adversary controls only one path. Within the considered model, there are several possible protection options. At one extreme, each primary path can be protected by two additional protection paths that are exclusively provisioned for it. This is a special case of our model. At the other extreme, one can consider provisioning protection paths that simultaneously protect all the primary paths. There also exist a host of intermediate strategies that may be less resource expensive. In this sense, our model captures a wide variety of protection options. However, the model does not capture scenarios where the uncoded unicast traffic travels over different primary paths. The model considers wired networks only and does not capture the characteristics of wireless networks.

Our work is a significant generalization of \cite{kamalR08}. We
assume the omniscient adversary model \cite{Jaggi08}, under which
the adversary has full knowledge of all details of the protocol
(encoding/decoding algorithms, coefficients, etc.) and has no
secrets hidden from her. An adversary changes data units on
several paths, which may be primary paths or protection paths. The
number of errors equals the number of paths the adversary attacks.
If multiple paths share one link and the adversary controls that
link, it is treated as multiple errors. Our schemes enable all
nodes to recover from $n_e$ errors, provided that $4n_e$
protection paths are shared by all the connections. More
generally, if there are $n_e$ adversarial errors and $n_f$
failures, a total of $4n_e + 2n_f$ protection paths are
sufficient. We emphasize that the number of protection paths only
depends on the number of errors and failures being protected
against and is independent of the number of unicast connections.
Simulation results show that if the number of primary paths is
large, the proposed protection scheme consumes less network
resources compared to the 2+1 protection scheme, where 2+1 means
that we use two additional paths to protect each primary
connection.


Section \ref{sec:net_encoding_protocol} introduces the network
model and our encoding protocol, which is a generalization of
\cite{kamalR08}. The error model is explained in Section
\ref{sec:error_model}. In Section \ref{sec:single error}, we
present the decoding algorithm and conditions when a single error
happens. Generalizations to multiple errors and combinations of
errors and failures are considered in Section
\ref{sec:multiple_errors} and Section \ref{sec:hybrid}. In Section
\ref{sec:result}, we briefly show how the optimal primary and
protection paths are provisioned by integer linear programming and
the simulation shows that our proposed approach saves network
resources. Section \ref{sec:conclusion} concludes the paper.

\section{Network model and encoding protocol}\label{sec:net_encoding_protocol}


Suppose that $2n$ nodes in the network establish $n$ bidirectional
unicast connections with the same capacity. These nodes are
partitioned into two disjoint sets $\calS$ and $\calT$ such that
each node in $\calS$ connects to one node in $\calT$. The $n$
connections are labeled by numbers $1,\ldots,n$ and the nodes
participating in the $i$th connection are given index $i$, i.e.,
$S_i$ and $T_i$. Each connection contains one bidirectional
primary path $S_i-T_i$. $S_i$ and $T_i$ send data units they want
to transmit onto the primary path. The data unit sent from $S_i$
to $T_i$ (from $T_i$ to $S_i$ respectively) on the primary path is denoted by
$d_i$ ($u_i$ respectively). The data unit received on the primary path by $T_i$
($S_i$ respectively) is denoted by $\dihat$ ($\uihat$ respectively).

A protection path $\bfP$ is a bidirectional path going through all
$2n$ end nodes of the $n$ connections. It has the same capacity as
the primary paths and consists of two unidirectional paths $\bfS$
and $\bfT$ in opposite directions. $M$ protection paths are used
and we assume that there are enough resources in the network so
that these protection paths can always be found and provisioned.
In this paper we mainly focus on the case where all protection
paths pass through all $2n$ end nodes of the connections, see Fig.
\ref{fig:example-enumeration} for an example, and they are denoted
by $\bfP^{(1)},\ldots,\bfP^{(M)}$. The order in which the
protection paths pass through the end nodes does not matter. The
more general case where different primary path connections are
protected by different protection paths will be discussed in
Section \ref{subsec:gen_topo}. All operations are over the finite
field $GF(q)$, $q=2^r$, where $r$ is the length of the data unit
in bits. Frequently used notations in this paper are summarized in
Table \ref{tab:Notations}.

\begin{table}[h]
\begin{center}
\caption{\label{tab:Notations} Frequently used notations in this
paper.}
\begin{tabular}{|c|l|}
\hline Notation & Meaning \\ \hline $n$ & The number of primary
connections \\ \hline $M$ & The number of protection paths \\
\hline $S_i,T_i$ & The end nodes of the $i^{th}$ primary
connection \\ \hline $d_i, u_i$ & The data unit sent by $S_i$,
$T_i$ respectively \\ \hline $\dihat,\uihat$ & The data unit
received by $T_i$, $S_i$ respectively \\ \hline  \multirow{2}{*} {
$\alpha_i^{(k)}, \beta_i^{(k)}$} & The encoding coefficients for
the $i^{th}$ primary \\ &  connection on the $k^{th}$ protection path \\
\hline $n_e, n_f$ & The number of errors and failures in the network\\
\hline \multirow{2}{*} {$n_c,n_p$} & The number of errors on the
primary paths \\ & and the protection paths respectively \\ \hline
$e_{d_i},e_{u_i}$ & The error values of $d_i, u_i$ respectively \\
\hline

\end{tabular}
\end{center}
\end{table}

The system works in rounds. Time is assumed to be slotted. Each
data unit is assigned a round number. In each round a new data
unit $d_i$ or $u_i$ is transmitted by node $S_i$ or $T_i$ on its
primary path. In addition, it also transmits an appropriately
encoded data unit in each direction on the protection path. The
encoding operation is executed by each node in $\calS$ and
$\calT$, where all nodes have sufficiently large buffers. The
encoding and decoding operations only take place between data
units of the same round. When a node is transmitting and receiving
data units of certain round on the primary path, it is receiving
data units of earlier rounds from the protection paths. The nodes
use the large, though bounded-size buffer to store the transmitted
and received data units for encoding and decoding. Once the
encoding and decoding for a certain round is done, the data units
of that round can be removed from the buffer. Overall, this
ensures that the protocol works even when there is no explicit
time synchronization between the transmissions.

Each connection $S_i-T_i$ has $2M$ encoding coefficients:
$\alpha_i^{(1)},\ldots,\alpha_i^{(M)},\beta_i^{(1)},\ldots,\beta_i^{(M)}$,
where $\alpha_i^{(k)}$ and $\beta_i^{(k)}$ are used for encoding
on protection path $\bfP^{(k)}$. Each protection path uses the
same protocol but different coefficients in general. The
coefficients are assumed to be known by the end nodes before the
transmission. We specify the protocol for protection path
$\bfP^{(k)}$, which consists of two unidirectional paths
$\bfS^{(k)}$ and $\bfT^{(k)}$. We first define the following
notations.
\begin{itemize}
\item $\sigma(S_i)/\sigma(T_i)$: the next node downstream from
$S_i$ (respectively $T_i$) on $\bfS^{(k)}$.
$\sigma^{-1}(S_i)/\sigma^{-1}(T_i)$: the next node upstream from
$S_i$ (respectively $T_i$) on $\bfS^{(k)}$ (see example in Fig.
\ref{fig:example-enumeration}). \item $\tau(S_i)/\tau(T_i)$: the
next node downstream from $S_i$ (respectively $T_i$) on
$\bfT^{(k)}$. $\tau^{-1}(S_i)/\tau^{-1}(T_i)$: the next node
upstream from $S_i$ (respectively $T_i$) on $\bfT^{(k)}$ (see
example in Fig. \ref{fig:example-enumeration}).
\end{itemize}
Each node transmits to its downstream node, the sum of the data
units from its upstream node and a linear combination of the data
units it has, on each unidirectional protection path. Consider the
$k^{th}$ protection path $\bfP^{(k)}$, denote the data unit
transmitted on link $e \in \bfS^{(k)}$ ($e \in \bfT^{(k)}$) by
$\bfS_e$ ($\bfT_e$). Node $S_i$ knows $d_i$,$\uihat$, and $T_i$
knows $u_i$, $\dihat$. The encoding operations are as
follows.\vspace{-2mm} \bean \bfS_{S_i \rightarrow \sigma(S_i)} &=&
\bfS_{\sigma^{-1}(S_i) \rightarrow
S_i} + \alpha_i^{(k)} d_i + \beta_i^{(k)} \uihat,\\
\bfT_{S_i \rightarrow \tau(S_i)} &=& \bfT_{\tau^{-1}(S_i)
\rightarrow S_i} + \alpha_i^{(k)} d_i + \beta_i^{(k)} \uihat,\\
 \bfS_{T_i \rightarrow \sigma(T_i)} &=&
\bfS_{\sigma^{-1}(T_i) \rightarrow T_i} +  \alpha_i^{(k)} \dihat +
\beta_i^{(k)} u_i, \text{and}\\
\bfT_{T_i \rightarrow \tau(T_i)} &=& \bfT_{\tau^{-1}(T_i)
\rightarrow T_i} + \alpha_i^{(k)} \dihat + \beta_i^{(k)} u_i.
\vspace{-2mm} \eean We focus our discussion on node $T_i$. Once
node $T_i$ receives data units over both $\bfS^{(k)}$ and
$\bfT^{(k)}$ it adds these data units. Denote the sum as
$P^{(k)}$\footnote{The values of $P^{(k)}$ are different at
different end nodes. Here we focus our discussion on node $T_i$.
To keep the notation simple, we use $P^{(k)}$ instead of
$P^{(k)}_{T_i}$} . $T_i$ gets two values
$\bfS_{\sigma^{-1}(T_i)\rightarrow T_i}$ and $\bfT_{\tau^{-1}(T_i)
\rightarrow T_i}$ from $\bfP^{(k)}$, $P^{(k)}$ equals
\beq\label{eq:pk}
 \bfS_{\sigma^{-1}(T_i)
\rightarrow T_i} + \bfT_{\tau^{-1}(T_i) \rightarrow T_i}
=\sum_{l:S_l \in \calS} \alpha_l^{(k)} d_l + \sum_{l:T_l \in
\calT\backslash\{T_i\}} \beta_l^{(k)} u_l + \sum_{l:S_l \in \calS}
\beta_l^{(k)} \hat{u}_l + \sum_{l:T_l \in \calT\backslash\{T_i\}}
\alpha_l^{(k)} \hat{d}_l. \eeq  In the absence of any errors,
$d_l=\hat{d}_l$, $u_l=\hat{u}_l$ for all $l$, most terms cancel
out because the addition operations are performed over an
extension field of the binary field and $P^{(k)} = \alpha_i^{(k)}
d_i + \beta_i^{(k)} \uihat$. Similar expressions can be derived
for the other end nodes. See Fig. \ref{fig:example-enumeration}
for an example of the encoding protocol.

\section{Error Model}\label{sec:error_model}\vspace{-1mm}
If the adversary changes data units on one (primary or protection)
path, \textit{an error} happens. If the adversary controls a link
through which multiple paths pass, or the adversary controls
several links, multiple errors occur. We assume that the adversary
knows the communication protocols described above, including the
encoding/decoding function and encoding coefficients. There are no
secrets hidden from her. If a primary or protection path is under
the control of an adversary,
she can arbitrarily change the data units in each direction on that path. 
If $d_i\neq \dihat$ or $u_i\neq \uihat$ (or both), we say that
there is \textit{an error on primary path } $S_i-T_i$ with
\textit{error values} $e_{d_i}=\di+\dihat$ and
$e_{u_i}=\ui+\uihat$. As for protection path error, although the
error is bidirectional, we shall see that each node will see only
one error due to the nature of the encoding protocol. In fact,
even multiple errors on the same protection path can be shown to
only have an aggregate effect as one error at one node. This is
because from one protection path, only the sum ($P^{(k)}$) of data
units from two directions is used in decoding at a node. If this
data unit is changed due to several errors, it can be modeled as
one variable $e_{p_k}$ at the node. However, different nodes will
have different values of $e_{p_k}$ in general. If there is a
primary path  failure (as opposed to error) on $S_i-T_i$, we have
$\dihat=\uihat=0$. i.e. failures are not adversarial. If a
protection path fails, it becomes useless and the end nodes ignore
the data units on that path. All nodes know the locations of
failures but do not know the locations of errors.

When there are errors in the network, the error terms will not
cancel out in \eqref{eq:pk} and $T_i$ obtains
$P^{(k)} = \alpha_i^{(k)}d_i + \beta_i^{(k)} (u_i+e_{u_i}) + \sum_{l\in I_{\backslash i}} (\alpha_l^{(k)}e_{d_l} + \beta_l^{(k)} e_{u_l}) + e_{p_k} $ 
on  protection path $\bfP^{(k)}$, where $I_{\backslash i} = \{
1,\ldots,n\}\backslash \{i\}$, the index set excluding $i$, and
$e_{p_k}$ is the error on protection path $\bfP^{(k)}$ seen by
$T_i$. Note that since $T_i$ knows $u_i$, we can subtract it from
this equation. Together with the data unit $P_m$ from the primary
path, $T_i$ has the following data units.\vspace{-2mm}
\bea\label{eq:tieqs}
P_m = \hat{d}_i&= &d_i + e_{d_i},  \label{eq:tieq1}\\
P^{(k)'} = P^{(k)} - \beta_i^{(k)} u_i & = &\alpha_i^{(k)}d_i +
\beta_i^{(k)} e_{u_i} + \sum_{l\in I_{\backslash i}}
(\alpha_l^{(k)}e_{d_l} + \beta_l^{(k)} e_{u_l}) + e_{p_k}, k =1,
\ldots, M \label{eq:tieq2}\vspace{-2mm} \eea We multiply
\eqref{eq:tieq1} by $\alpha_i^{(k)}$ and add to the $k^{th}$
equation in \eqref{eq:tieq2} to obtain \vspace{-1mm}
\beq\label{eq:error_model_single} \sum_{l=1}^{n} (\alpha_l^{(k)}
e_{d_l} + \beta_l^{(k)} e_{u_l}) + e_{p_k}= \alpha_i^{(k)} P_m +
P^{(k)'}, k = 1,\ldots,M. \eeq This can be represented in matrix
form as \vspace{-2mm}

\beq\label{eq:gen_error_pr} \left[\begin{array}{cccccccccccc}
\alpha_1^{(1)} & \beta_1^{(1)}  & \cdots & \alpha_n^{(1)} & \beta_n^{(1)} & 1 & 0 & \cdots & 0   \\
\alpha_1^{(2)} & \beta_1^{(2)}  & \cdots & \alpha_n^{(2)} & \beta_n^{(2)} & 0 & 1 & \cdots & 0 \\
\vdots & \vdots & \vdots & \vdots & \vdots & \vdots & \vdots & \vdots & \vdots\\
\alpha_1^{(M)} & \beta_1^{(M)}  & \cdots & \alpha_n^{(M)} &
\beta_n^{(M)} & 0 & 0 & \cdots & 1 \end{array} \right]  E =
P_{syn},
\eeq where the length-$(2n+M)$ vector $E = [e_{d_1}, e_{u_1},
\ldots, e_{d_n}, e_{u_n}, e_{p_1},\ldots, e_{p_M}]^T$ and the
length-$M$ vector $P_{syn}=[\alpha_i^{(1)}P_m+P^{(1)'},
\alpha_i^{(2)}P_m+P^{(2)'},\ldots,\alpha_i^{(M)}P_m+P^{(M)'}]^T$.
Analogous to classical coding theory, we call $P_{syn}$ the {\it
syndrome} available at the decoder. Denote the $M\times (2n+M)$
coefficient matrix of \eqref{eq:gen_error_pr} as $H_{ext}$, and
denote the first $2n$ columns of $H_{ext}$ as a matrix $H =
[\bfv_1, \bfv_2, \ldots, \bfv_{2n}]$, where $\mathbf{v}_j$ is the
$j^{th}$ column of $H$. Then $\bfv_{2i-1}, \bfv_{2i} $ are the
columns consisting of encoding coefficients $\alpha_i$'s and
$\beta_i$'s for the connection $S_i-T_i$. The last $M$ columns of
$H_{ext}$ form an identity matrix $I_{M\times M}$ and can be
denoted column by column as $[\bfv^p_1, \ldots, \bfv^p_M]$. Note
that $T_i$ knows $H$ and $P_{syn}$ and shall attempt to decode
$d_i$ even in the presence of the errors. Node $S_i$ gets very
similar equations to those at $T_i$. Thus we will focus our
discussion on $T_i$. Each end node uses the same decoding
algorithm and works individually without cooperation and without
synchronization.

%

\section{Recovery from single error}\label{sec:single error} 

In this section, we focus on the case when there is only one error in the network. We first present the decoding algorithm and then prove its correctness under appropriate conditions.
\vspace{-3mm}
\subsection{Decoding algorithm at node $T_i$ ($S_i$ operates similarly)}
\vspace{-2mm}

\begin{enumerate}
\item Attempt to solve the following system of equations 
\beq\label{eq:t1error1} [\bfv_{2i-1} \bfv_{2i}]
\left[\begin{array}{c}
e_{d_i}\\
e_{u_i}\end{array}\right]= P_{syn} \vspace{-2mm} \eeq \item If
\eqref{eq:t1error1} has a solution $(e_{d_i},e_{u_i})$, compute
$d_i=P_m+e_{d_i}$, otherwise, $d_i=P_m$
\end{enumerate}

We show below that this algorithm works when the error happens on
a primary path or on one of the protection paths. \vspace{-3mm}
\subsection{Condition for one primary path error correction}\vspace{-2mm} 
In this subsection, we consider primary path error only. Define an
\textit{error pattern} to be the two columns in $H$ corresponding
to the erroneous primary path. If the error happens on $S_i-T_i$,
the error pattern is $\{\bfv_{2i-1},\bfv_{2i}\}$. An \textit{error
value vector} corresponding to an error pattern is obtained by
letting the error values corresponding to other $n-1$ primary
paths to be zero. The error value vector corresponding to error
pattern $\{\bfv_{2i-1},\bfv_{2i}\}$ is the length-$2n$ vector
$E_i=[0,\ldots,e_{d_i},e_{u_i},\ldots,0]^T$.
Assume that $e_{d_i}$'s and $e_{u_i}$'s are not all zero. The case when all of them are zero is trivial because it implies that no error happens. 

\begin{theorem}\label{theo:1error_big}
Suppose there is at most one error on a primary path. The decoding algorithm outputs the correct data unit at every node if and only if the vectors in the set $\{\bfv_{2i-1},\bfv_{2i},\bfv_{2j-1},\bfv_{2j}\}$\footnote{In fact, it can be viewed as the error pattern when $S_i-T_i, S_j-T_j$ are in error.} for all $i,j = 1,\ldots,n, i \neq j$ are linearly independent. 
\end{theorem}

\emph{Proof:} First assume that the vectors in the sets
$\{\bfv_{2i-1},\bfv_{2i},\bfv_{2j-1},\bfv_{2j}\}$ are linearly
independent. Let $E_a$ and $E_b$ be error value vectors
corresponding to errors happening on different primary paths
$S_a-T_a$ and $S_b-T_b$ respectively. Suppose there exist $E_a$
and $E_b$ such that $HE_a=HE_b$, i.e., $H(E_a+E_b)=0$. Note that
the vector $(E_a+E_b)$ has at most four error values
$[e_{d_a},e_{u_a},e_{d_b},e_{u_b}]$ which are not all zero and
such that
$[\begin{array}{ccccc}
\bfv_{2a-1},\bfv_{2a},\bfv_{2b-1},\bfv_{2b}\end{array}][e_{d_a},e_{u_a},e_{d_b},e_{u_b}]^T
= \mathbf{0}$.
This implies $\{\bfv_{2a-1},\bfv_{2a},\bfv_{2b-1},\bfv_{2b}\}$ are
linearly dependent, which is a contradiction. Therefore, under our
condition that $\{\bfv_{2i-1},\bfv_{2i},\bfv_{2j-1},\bfv_{2j}\}$
for all $i,j = 1,\ldots,n, i \neq j$ are linearly independent,
there does not exist $E_a,E_b$ such that $HE_a=HE_b$. This means
that if we try to solve the system of linear equations according
to every possible error value vectors $E_1,\ldots, E_n$, it either
has no solution or its solution is the actual error in the
network. The node $T_i$ is only interested in $d_i$, in our
decoding algorithm, it tries to solve the equations
\eqref{eq:t1error1} according to the error value vector $E_i$. If
it has a solution, the error happens on $S_i-T_i$. The matrix
$[\bfv_{2i-1},\bfv_{2i}]$ has rank two, so equations
\eqref{eq:t1error1} have unique solution for $e_{d_1}$.
$d_i=P_m+e_{d_i}$ gives decoded $d_i$. If \eqref{eq:t1error1} does
not have solution, the error is not on $S_i-T_i$. $T_i$ simply
picks up $d_i=P_m$ from the primary path $S_i-T_i$.

Conversely, suppose that a vector set
$\{\bfv_{2i_1-1},\bfv_{2i_1},\bfv_{2j_1-1},\bfv_{2j_1}\}$ is
linearly dependent. There exist $E_{i_1}$ and $E_{j_1}$ such that
$HE_{i_1}=HE_{j_1}$. Both equations $HE_{i_1}=P_{syn}$ and
$HE_{j_1}=P_{syn}$ have solution. Suppose the error in fact
happens on $S_{j_1}-T_{j_1}$, the decoder at $T_{i_1}$ can also
find a solution to $HE_{i_1}=P_{syn}$ and use the solution to
compute $d_i$. This leads to decoding error. \endproof

If there is no error in the network, $P_{syn}=0$ and solving
\eqref{eq:t1error1} gives $e_{d_i}=e_{u_i}=0$.
In order to make $\{\bfv_{2i-1},\bfv_{2i},\bfv_{2j-1},\bfv_{2j}\}$
independent, we need the length of vectors to be at least four,
i.e., $M\geq 4$. In fact, we shall see that several coefficient
assignment strategies ensure that four protection paths are
sufficient to make the condition hold for $\forall i,j =
1,\ldots,n, i\neq j$. The condition in Theorem
\ref{theo:1error_big} can be stated as all $M\times M$ ($4\times
4$) matrices of the form \vspace{-2mm}
\beq\label{eq:1error_fullrank}
[\bfv_{2i-1},\bfv_{2i},\bfv_{2j-1},\bfv_{2j}],i,j = 1,\ldots,n,
i<j \vspace{-2mm} \eeq have full rank.

\vspace{-3mm}
\subsection{Coefficient assignment methods}\label{subsec:simple_scheme} \vspace{-1mm}
We shall introduce several ways to assign encoding coefficients,
so that \eqref{eq:1error_fullrank} has full rank. Later we will
see these schemes also work when protection path error is
possible.

\begin{list}{}{\leftmargin=0.0cm \labelwidth=0cm \labelsep = 0cm}

\item[$(1)$ ] {\it A simple scheme of coefficient assignment and
implementation.}
Choose $n$ non-zero distinct elements $\gamma_1,\ldots,\gamma_n$
from $GF(q)$. For all $i=1,\ldots,n$, $\alpha_i^{(1)} = 1$,
$\alpha_i^{(2)} = \gamma_i$, $\beta_i^{(3)} = 1$, $\beta_i^{(4)} =
\gamma_i$ and all other coefficients are zero.
It can be shown by performing Gaussian elimination that the matrix
\eqref{eq:1error_fullrank} has full rank as long as $\gamma$'s are
distinct. The minimum field size needed is $q>n$.

Consider decoding at node $T_i$, Table \ref{tab:simple_coeff} is a
summary of the data units $P_m, P_{syn}$ that $T_1$ gets from
primary path and protection paths under different cases.
$P_{syn}^{(k)}$ is the $k^{th}$ component of $P_{syn}$. The
decoding is done as follows. If $P_{syn}^{(1)}$ and
$P_{syn}^{(2)}$ are both zero, then $e_{d_l} = 0, \forall l$, $T_i$
simply pick $d_i = P_m$. If $P_{syn}^{(1)}$ and $P_{syn}^{(2)}$
are both non-zero, $T_i$ computes $S=P_{syn}^{(2)} \times
(P_{syn}^{(1)})^{-1}$. If $S=\gamma_i$, the error happens on
$S_i-T_i$ and the error value is $e_{d_i} = P_{syn}^{(1)}$, then
$d_i = P_m + e_{d_i}$. If $S = \gamma_x$, the error happens on
$S_x-T_x, x\neq i$, then $T_i$ picks up $d_i = P_m$.

Note that we only used $P_m, P_{syn}^{(1)},P_{syn}^{(2)}$ to
decode $d_i$ at $T_i$. However, we cannot remove paths
$\bfP^{(3)},\bfP^{(4)}$ because at $S_i$ we should use $P_m,
P_{syn}^{(3)},P_{syn}^{(4)}$ to decode.

\begin{table}[h]
\begin{center}
\caption{\label{tab:simple_coeff} Data obtained by $T_i$ under the
simple coefficient assignment.}
\begin{tabular}{|c|c|c|c|}
\hline
 & No error &Error on $S_i-T_i$ & Error on $S_x-T_x,i\neq x$\\ \hline
$P_m$ &  $d_i$ &  $d_i+e_{d_i}$ &  $d_i$\\
$P_{syn}^{(1)}$&  $0$ &  $e_{d_i}$ &  $e_{d_x}$\\
$P_{syn}^{(2)}$&  $0$ &  $\gamma_i e_{d_i}$ &  $\gamma_x e_{d_x}$\\
$P_{syn}^{(3)}$&  0 &  $e_{u_i}$ & $ e_{u_x}$\\
$P_{syn}^{(4)}$&  0 &  $\gamma_i e_{u_i}$ &  $\gamma_x e_{u_x}$\\
\hline
\end{tabular}
\end{center}
\end{table}

\item[$(2)$ ] {\it Vandermonde matrix.} The second way is to
choose $2n$ distinct elements from
$GF(q):\gamma_{\alpha_1},\gamma_{\beta_1},\ldots,\gamma_{\alpha_n},\gamma_{\beta_n}$
and let encoding coefficients to be
$\alpha_i^{(k)}=\gamma_{\alpha_i}^{k-1},\beta_i^{(k)}=\gamma_{\beta_i}^{k-1}$.
The matrix in equation \eqref{eq:1error_fullrank} becomes a
Vandermonde matrix and has full rank.
\item[$(3)$ ] {\it Random choice.} Besides the structured matrices
above, choosing coefficients at random from a large field also
works with high probability due to the following claim.
\begin{claim}\label{clm:1err_ran}
When all coefficients are randomly, independently and uniformly
chosen from $GF(q)$, for given $i$ and $j$, the probability that
$\{\bfv_{2i-1},\bfv_{2i},\bfv_{2j-1},\bfv_{2j}\}$ are linearly
independent is $p_1=(1-1/q^3)(1-1/q^2)(1-1/q)$.
\end{claim}
\emph{Proof:} Suppose we have chosen $\bfv_{2i-1}$, the
probability that $\bfv_{2i}$ is not in the span of $\bfv_{2i-1}$
is $(1-q/q^4)$. The probability that $\bfv_{2j-1}$ is not in the
span of $\{\bfv_{2i-1},\bfv_{2i}\}$ is $(1-q^2/q^4)$. The
probability that $\bfv_{2j}$ is not in the span of
$\{\bfv_{2i-1},\bfv_{2i},\bfv_{2j-1}\}$ is $(1-q^3/q^4)$. Since
the coefficients are chosen independently, the probability that
four vectors are linearly independent is the product $p_1$, which
approaches 1 when $q$ is large.\endproof In
\eqref{eq:1error_fullrank} we require ${n \choose 2}$ matrices to
have full rank. By union bound, the probability that the linear
independence condition in Theorem \ref{theo:1error_big} holds is
at least $1-(1-p_1){n \choose 2}$, which is close to 1 when $q$ is
large. In practice, before all the transmission, we could generate
the coefficients randomly until they satisfy the condition in
Theorem \ref{theo:1error_big}. Then, transmit those coefficients
to all the end nodes in the network. During the actual
transmission of the data units, the encoding coefficients do not
change.

\end{list}
\vspace{-2mm}
\subsection{Taking protection path error into account} \vspace{-2mm} 
In this subsection, we take protection path errors into account.
The error (assume one error in this section) can happen either on
one primary path or one protection path. Besides $n$ error value
vectors $E_1,\ldots,E_n$, we have $M$ more error value vectors for
the protection path error:
$[\mathbf{0}|e_{p_1},0,\ldots,0]^T,\ldots,[\mathbf{0}|0,0,\ldots,e_{p_M}]^T$,
where $\mathbf{0}$ denote an all-zero vector of length $2n$.
Denote them by $E_{p_1},\ldots, E_{p_M}$. Using a similar idea to
Theorem \ref{theo:1error_big}, we have the following:
\begin{theorem}\label{theo:1error_big_pro}
If there is one error on one primary path or protection path, the decoding algorithm works for every node if and only if vectors in the sets 
\vspace{-2mm} \bea
\{\bfv_{2i-1},\bfv_{2i},\bfv_{2j-1},\bfv_{2j}\}, i,j = 1,\ldots,n,i\neq j \label{eq:pro_c1}\\
\{\bfv_{2i-1},\bfv_{2i},\bfv^p_l\},i=1,\ldots,n, l=1,\ldots,M
\label{eq:pro_c2} \eea \vspace{-2mm} are linearly independent.
Note that $\bfv^p_l$ is the $l^{th}$ column in $I_{M\times M}$ in
\eqref{eq:gen_error_pr}.
\end{theorem}

In fact, $M=4$ suffices and the three coefficient assignment
methods we described in the previous subsection work in this case.
The simple coefficient assignment strategy in Section
\ref{subsec:simple_scheme}(1) enables vector sets
\eqref{eq:pro_c1} and \eqref{eq:pro_c2} to be independent. The
protection path error makes exact one component of $P_{syn}$ to be
nonzero. If $T_i$ detects $P_{syn}$ has only one nonzero entry, it
can just pick up the data unit from the primary path since the
only error is on the protection path.

In order to see that Vandermonde matrix also works, we shall show
that the vector sets \eqref{eq:pro_c2} are linearly independent.
Suppose that they are linearly dependent. Since
$\bfv_{2i-1},\bfv_{2i}$ are linearly independent, there exist $a$
and $b$ such that (take $\bfv^p_1$ for example): $a \bfv_{2i-1} +
b \bfv_{2i} = \bfv^p_1$.
This means $a[\gamma_{\alpha_i} \gamma_{\alpha_i}^2]^T+b[
\gamma_{\beta_i} \gamma_{\beta_i}^2]^T=\mathbf{0}$. However, this
is impossible since \beqno
\det \left[\begin{array}{cc} \gamma_{\alpha_i} & \gamma_{\beta_i} \\
\gamma_{\alpha_i}^2& \gamma_{\beta_i}^2 \end{array} \right]\neq 0.
\eeqno

Therefore,  $\{\bfv_{2i-1},\bfv_{2i},\bfv^p_1\}$ are linearly
independent. A similar argument holds for $\bfv^p_l$ when $l\neq
1$.

When the coefficients are randomly chosen from $GF(q)$, for given
$i$ and $l$, the probability that
$\{\bfv_{2i-1},\bfv_{2i},\bfv^p_l\}$ are linearly independent is
$p_2=(1-1/q^3)(1-1/q^2)$. Considering all vector sets in Theorem
\ref{theo:1error_big_pro}, the probability of successful decoding
at all nodes is at least $1-(1-p_1){n \choose 2}-(1-p_2)nM$, which
approaches 1 when $q$ is large.
\vspace{-2mm}
\subsection{Remark}\vspace{-2mm}
We can compare our results with classical results in coding
theory. In classical coding theory, in the presence of two
adversarial errors, we need a code with minimum distance at least
five for correct decoding. This means that to transmit one symbol
of information, we need to transmit a codeword with at least five
symbols. In our problem, each connection has a total of five paths
(one primary and four protection). A single error on a
bidirectional primary path induces two errors, one in each
direction. Therefore in an approximate sense we are using almost
the optimal number of protection paths. However, a proof of this
statement seems to be hard to arrive at. It is important to note
that the protection paths are shared so the cost of protection per
primary path connection is small. \vspace{-2mm}
\subsection{The case when the primary paths are protected by different protection paths}\label{subsec:gen_topo}\vspace{-2mm}
If the primary paths are protected by different protection paths,
the models are similar. Specifically, consider node $T_i$ and it
is protected by the protection path $\bfP_k$, if we denote the set
of primary paths protected by protection path $\bfP_k$ by
$N(\bfP_k)\subseteq \{1,\ldots,n\}$, the equation obtained from
protection path $\bfP_k$ by $T_i$ is similar to
\eqref{eq:error_model_single}: $\sum_{l\in
N(\bfP_k)}(\alpha_l^{(k)} e_{d_l} + \beta_l^{(k)} e_{u_l}) +
e_{p_k} = \alpha_i^{(k)} P_m + P^{(k)'}. $ Now, $T_i$ obtains
$M_i$ equations, where $M_i$ is the number of protection paths
protecting connection $S_i-T_i$. The system of equations it gets
is similar to \eqref{eq:gen_error_pr}, but the $M_i\times 2n$
coefficient matrix $H$ may contain zeros induced by the network
topology. If connection $S_l-T_l$ is not protected by $\bfP_k$,
the corresponding two terms in the $k$th row are zero. The
identity matrix in $H_{ext}$ is $I_{M_i\times M_i}$. The models
are similar to the case when all connections are protected by the
same protection paths and the decoding algorithms and conditions
in Theorem \ref{theo:1error_big} and \ref{theo:1error_big_pro}
still work.

The difference comes from the coefficient assignment. $H$ may
contain some zeros depending on the topology. In order to make
\eqref{eq:pro_c1},\eqref{eq:pro_c2} to be linearly independent, we
can use the method of matrix completion \cite{harvey05}. We view
the encoding coefficients in $H$ as indeterminates to be decided.
The matrices we require to have full rank are a collection ${\cal
C}_{H}$ of submatrices of $H_{ext}$, where ${\cal C}_{H}$ depends
on the network topology. Each matrix in ${\cal C}_{H}$ consists of
some indeterminates and possibly some zeros due to the topological
constraints and ones coming from the last $M_1$ columns of
$H_{ext}$. The problem of choosing encoding coefficients can be
solved by matrix completion. A simultaneous max-rank completion of
${\cal C}_{H}$ is an assignment of values from $GF(q)$ to the
indeterminates that preserves the rank of all matrices in ${\cal
C}_{H}$. After completion, each matrix will have the maximum
possible rank. Note that if $H$ contains too many zeros, it may be
not possible to make the matrices to have the required rank when
$M_i = 4$. Thus, $M_i = 4$ is a necessary but not in general
sufficient condition for successful recovery.
It is known that choosing the indeterminates at random from a
sufficiently large field can solve the matrix completion problem
with high probability \cite{lovasz79}.  Hence, we can choose
encoding coefficients randomly from a large field. It is clear
therefore that the general case can be treated conceptually in a
similar manner to what we discussed earlier. Thus, we shall mainly
focus on the case when the protection paths protect all the
primary paths.

\vspace{-1.5mm}
\section{Recovery from multiple errors}
\label{sec:multiple_errors}

Our analysis can be generalized to multiple errors on primary and
protection paths. Assume that $n_c$ errors happen on primary paths
and $n_p=n_e-n_c$ errors happen on protection paths. As described
in Section \ref{sec:error_model}, a given primary path error
corresponds to two specific columns in $H_{ext}$ while a
protection path error corresponds to one specific column in
$H_{ext}$. Recall that we view $H_{ext}$ as a set of column
vectors : $\{\bfv_1,
\bfv_2,\ldots,\bfv_{2n-1},\bfv_{2n},\bfv^p_1,\bfv^p_2,\ldots,\bfv^p_M\}$.
An error pattern is specified by the subset of columns of
$H_{ext}$ corresponding to the paths in error.

\begin{definition}
A subset of columns of $H_{ext}$ denoted as $A(m_1,m_2)$ is {\it
an error pattern} with $m_1$ errors on primary paths
$\{c_1,\ldots,c_{m_1}\} \subseteq \{1,\ldots,n\}$ and $m_2$ errors
on protection paths $\{p_1,\ldots,p_{m_2}\} \subseteq
\{1,\ldots,M\}$ if it has the following form:
$A(m_1,m_2)=A_{c}(m_1)\cup A_{p}(m_2)$, where $A_{c}(m_1)
=\{\bfv_{2c_1-1},\bfv_{2c_1},~\ldots~,\bfv_{2c_{m_1}-1},\bfv_{2c_{m_1}}\}$,
$~c_i\in \{1,\ldots,n\}, \forall i=1,\ldots,m_1$ and $A_{p} (m_2)
=\{\bfv^p_{p_1},\ldots,\bfv^p_{p_{m_2}}\},p_i\in
\{1,\ldots,M\},\forall i=1,\ldots,m_2$.
\end{definition}
Note that $|A(m_1,m_2)|=2m_1+m_2$ and the set of columns in $H_{ext}$ can be expressed as $A(n,
M)$. Although our definition of error pattern is different from the conventional definition in classical coding theory, we shall find it helpful for the discussion of our algorithms.

We let $\mathbf{A}(m_1,m_2)$ denote the family of error patterns
with $m_1$ primary path errors and $m_2$ protection path errors
(for brevity, henceforth we refer to such errors as $(m_1, m_2)$
type errors).

\begin{definition} Define $\mathbf{A}(m_1,m_2)_i$, a subset of $\mathbf{A}(m_1,m_2)$, to be the family of $(m_1, m_2)$ type error
patterns such that each error pattern includes an error on primary
path $S_i - T_i$, i.e., $A(m_1, m_2) \in \mathbf{A}(m_1,m_2)_i$ if
and only if $ \{\bfv_{2i-1},\bfv_{2i}\} \subseteq A(m_1,m_2)$.
\end{definition}

Note that $|\mathbf{A}(m_1,m_2)|={n \choose m_1}{M \choose m_2}$
and $|\mathbf{A}(m_1,m_2)_i| = {n-1 \choose m_1-1}{M \choose
m_2}$. Denote the family of error patterns including an error on
$S_i-T_i$ with $n_e$ errors in total as:
$\mathbf{A_i}(n_e)=\cup_{n_c=1}^{n_e} \mathbf{A}(n_c, n_e-n_c)_i$.

Our definition of an error pattern has only specified the location
of the error but not the actual values.
An error value vector $E$ has the following form
:$[e_{d_1},e_{u_1},\ldots,e_{d_n},e_{u_n},e_{p_1},\ldots,e_{p_M}]^T$.
Each entry of the vector corresponds to one column in $H_{ext}$.
An error value vector $E$ corresponds to an error pattern
$A(m_1,m_2)$ if in $E$, the entries corresponding to $A(n,M)
\backslash A(m_1,m_2)$ are zero, while the other entries may be
non-zero and are indeterminates in the decoding algorithm. We are
now ready to present the decoding algorithm in the presence of
multiple errors. \vspace{-2mm}
\subsection{Multiple errors decoding algorithm at node $T_i$ ($S_i$ operates similarly)}
\label{subsec:multi_algo}\vspace{-2mm}
\begin{enumerate}
\item Try to solve the system of linear equations specified in
\eqref{eq:gen_error_pr} according to each error pattern in
$\mathbf{A_i}(n_e)$. This means for each error pattern in
$\mathbf{A_i}(n_e)$, replace $E$ in \eqref{eq:gen_error_pr} by the
error value vector, which contains the indeterminates,
corresponding to the error pattern. \item Suppose that the decoder
finds a solution to one of these system of equations. Compute $d_i
= P_m+e_{d_i}$, where $e_{d_i}$ is recovered as part of the
solution. If none of these systems of equations has a solution,
set $d_i=P_m$.
\end{enumerate}
This algorithm requires the enumeration of all error patterns in
$\mathbf{A_i}(n_e)$ and has high computational complexity
(exponential in the number of errors). In Section
\ref{su0bsec:RSefficient}, a low complexity polynomial-time
algorithm will be proposed under the assumption that the errors
only happen on the primary paths.

 \vspace{-3mm}
\subsection{Condition for error correction}
\vspace{-3mm}
\begin{theorem}\label{theo:multi_error}
Suppose that there are at most $n_e$ errors in the network (both
primary path error and protection path error are possible). The
result of the decoding algorithm is correct at every node if and
only if the column vectors in $A(m_1, m_2)$ are linearly
independent for all $A(m_1, m_2) \in \cup_{n_c, n_c' \in \{0,
\ldots, n_e\}} \mathbf{A}(n_c+n_c', 2n_e-(n_c+n_c'))$.
\end{theorem}

\emph{Proof:} First we shall show that under the stated condition,
the decoding algorithm works. Suppose $E_1$ and $E_2$ denote two
error value vectors corresponding to error patterns in
$\mathbf{A}(n_c,n_e-n_c)$ and  $\mathbf{A}(n_c',n_e-n_c')$
respectively and $E_1\neq E_2$. The linear independence condition
in the theorem implies that there do not exist $E_1$ and $E_2$
such that $HE_1=HE_2$. To see this, suppose there exist such $E_1$
and $E_2$, then, $HE_{sum}=0$, where $E_{sum}=E_1+E_2\neq 0$ has
at most $n_c+n_c'$ errors on primary paths and
$n_p+n_p'=2n_e-(n_c+n_c')$ errors on protection path. These errors
correspond to a member (which is a set of column vectors)
$A(n_c+n_c', 2n_e-(n_c+n_c'))\in \mathbf{A}(n_c+n_c',
2n_e-(n_c+n_c'))$. $HE_{sum}=0$ contradicts the linear
independence of the column vectors in $A(n_c+n_c',
2n_e-(n_c+n_c'))$. Thus, $E_1, E_2$ do not exist for $HE_1=HE_2$.
This means that if a decoder tries to solve every system of linear
equations according to every possible error patterns with $n_e$
errors, it either gets no solution, or gets the same solution for
multiple solvable systems of linear equations. A decoder at $T_i$
is only interested in error patterns in $\mathbf{A_i}(n_e)$. If in
step 1 it finds a solution $E$ for one system of equation,
$e_{d_i}$ in $E$ is the actual error value for $d_i$ and
$d_i=P_m+e_{d_i}$, otherwise, no error happens on $S_i-T_i$.

Conversely, if there exist some $n_c, n_c'$ such that some member
in $\mathbf{A}(n_c+n_c', 2n_e-(n_c+n_c'))$ is linearly dependent,
there exist $E_1'$ and $E_2'$ such that $HE_1'=HE_2'$ and
$E_1'\neq E_2'$. This implies that there exists an $i_1$ such that
either $e_{d_{i_1}}$ or $e_{u_{i_1}}$ is different. At node
$T_{i_1}$ or $S_{i_1}$, the decoder has no way to distinguish
which one is the actual error value vector and the decoding fails.
\endproof

The above condition is equivalent to the fact that all vector sets
$A(m_1,m_2) \in\cup_{m\in\{0,\ldots,2n_e\}}\mathbf{A}(m, 2n_e-m)$
are linearly independent. $|A(m,2n_e-m)|=2n_e+m$ and its maximum
is $4n_e$. Thus, the length of the vectors should be at least
$4n_e$. In fact, $M=4n_e$ is sufficient under random chosen
coefficients. Suppose that the coefficients are randomly and
uniformly chosen from $GF(q)$. For a fixed $m$, the probability
that $A(m,2n_e-m)=A_{c}(m)\cup A_{p}(2n_e-m)$ is linearly
independent is $p_1(m)=\prod_{i=0}^{2m-1}(1-q^{2n_e-m+i}/q^M)$.
Considering all members in $\mathbf{A}(m, 2n_e-m)$ and all values
of $m$, by union bound, the probability for successful decoding is
at least $1-\sum_{m=0}^{2n_e}(1-p_1(m)){n \choose m}{M \choose
2n_e-m}$, which approaches 1 when $q$ is large. \vspace{-2mm}
\subsection{Reed-Solomon like efficient decoding for primary path error only
case}\label{su0bsec:RSefficient} \vspace{-2mm} If the errors only
happen on primary paths, the condition in Theorem
\ref{theo:multi_error} becomes that each member of
$\mathbf{A}(2n_e,0)$ is linearly independent. We can choose $H$ so
that $H_{ij} = (\alpha^i)^{j-1}$, where $\alpha$ is the primitive
element over $GF(q)$, with $q>2n$. This is a parity check matrix
of a $(2n, 2n-M)$ Reed-Solomon code. Denote it by $H_{RS}$. Any
$M$ ($M=4n_e$) columns of $H_{RS}$ are linearly independent and
satisfies the condition in Theorem \ref{theo:multi_error}. Thus,
\eqref{eq:gen_error_pr} becomes
$H_{RS}[e_{d_1},e_{u_1},\ldots,e_{d_n},e_{u_n}]^T=P_{syn}$, in
which $H_{RS}$ and $P_{syn}$ are known by every node. The decoding
problem becomes to find an error pattern with at most $n_e$ errors
and the corresponding error value vector. Note that in fact there
are $2n_e$ error values to be decided. This problem can be viewed
as RS hard decision decoding problem while the number of errors is
bounded by $2n_e$. $P_{syn}$ can be viewed as the
\textit{syndrome} of a received message.
We can apply Berlekamp-Massey algorithm (BMA) for decoding. It is
an efficient polynomial time algorithm, while the proposed
algorithm in Section \ref{subsec:multi_algo} has exponential
complexity. Further details about RS codes and BMA can be found in
\cite{shulin}.

\vspace{-2mm}
\section{Recovery from a combination of errors and failures}\label{sec:hybrid} 

We now consider a combination of errors and failures on primary
and protection paths. Recall that when a primary path or a
protection path is in failure, then all the nodes are assumed to
be aware of the location of the failure. Assume that there are a
total of $n_f$ failures in the network, such that $n_{f_c}$
failures are on primary paths and $n_{f_p}=n_f-n_{f_c}$ failures
are on protection paths. If a protection path has a failure it is
basically useless and we remove the equation corresponding to it
in error model \eqref{eq:gen_error_pr}. Thus, we shall mainly work
with primary path failures and error model \eqref{eq:gen_error_pr}
will have $M'=M-n_{f_p}$ equations. In our error model,
when a primary path failure happens, $\dihat=0$ ($\uihat=0$
respectively). We can treat a primary path failure as a primary
path error with error value $e_{d_i}=d_i$ ($e_{u_i}=u_i$
respectively). In the failure-only case considered in
\cite{kamalR08}, $n_{f_c}$ protection paths are needed for
recovery from $n_{f_c}$ primary path failures. However, the
coefficients are chosen such that
$\alpha_i^{(k)}=\beta_i^{(k)},\forall i,k$,  which violates the
condition for error correction discussed before. Thus, we need
more paths when faced with a combination of errors and failures.


The decoding algorithm and condition in this case are very similar
to multiple error case. An important difference is that the
decoder knows the location of $n_f$ failures. To handle the case
of failures, we need to modify some definitions in Section
\ref{sec:multiple_errors}.

\begin{definition}
A subset of columns of $H$ denoted by $F(n_{f_c})$ is said to be
a {\it failure pattern} with $n_{f_c}$ failures on primary paths
$\{f_1,\ldots,f_{n_{f_c}}\}\subseteq \{1,\ldots,n\}$ if it has the
following form: $F(n_{f_c})=\{\bfv_{2f_1-1},\bfv_{2f_1},\ldots,
\bfv_{2{f_{n_{f_c}}}-1},\bfv_{2{f_{n_{f_c}}}}\}$,$f_i\in
\{1,\ldots,n\}$.
\end{definition}

\begin{definition}
An error/failure pattern with $m_1$ primary path errors, $m_2$
protection path errors and failure pattern $F(n_{f_c})$ is defined
as $A^F(m_1,m_2,F(n_{f_c}))=A(m_1,m_2)_{\backslash F(n_{f_c})}\cup
F(n_{f_c})$, where $A(m_1,m_2)_{\backslash F(n_{f_c})} \in
\bfA(m_1,m_2)$ and is such that $A(m_1,m_2)_{\backslash
F(n_{f_c})} \cap F(n_{f_c})=\emptyset$, i.e.,
$A(m_1,m_2)_{\backslash F(n_{f_c})}$ is a $(m_1,m_2)$ type error,
of which the primary path errors do not happen on failed paths in
$F(n_{f_c})$.
\end{definition}

We let $\bfA^F(m_1,m_2,F(n_{f_c}))$ denote the family of
error/failure patterns with $m_1$ primary path errors, $m_2$
protection path errors ($(m_1,m_2)$ type errors) and a fixed
failure pattern $F(n_{f_c})$.

\begin{definition} Define a subset of $\mathbf{A}^F(m_1,m_2,F(n_{f_c}))$, denoted as $\mathbf{A}^F(m_1,m_2,F(n_{f_c}))_i$ to be the family of error/failure
patterns such that each pattern includes an error or failure on
$S_i - T_i$, i.e., $A^F(m_1, m_2,F(n_{f_c})) \in
\mathbf{A}^F(m_1,m_2,F(n_{f_c}))_i$ if and only if
$\{\bfv_{2i-1},\bfv_{2i}\} \subseteq A^F(m_1,m_2,F(n_{f_c}))$.
\end{definition}

An error/failure value vector $E$ corresponds to an error/failure
pattern $A^F(m_1,m_2,F(n_{f_c}))$ if the entries corresponding to
$A(n,M) \backslash A^F(m_1,m_2,F(n_{f_c}))$ are zero, while the
other entries may be non-zero. \vspace{-3mm}
\subsection{Decoding algorithm at node $T_i$ for combined failures and errors ($S_i$ operates similarly) }
\vspace{-3mm}
\begin{enumerate}
\item Note that $T_i$ knows the failure pattern for all primary
paths $F(n_{f_c})$. It tries to solve equations of
\eqref{eq:gen_error_pr} form according to each error/failure
pattern in $\cup_{n_c=1}^{n_e} \mathbf{A}^F(n_c,
n_e-n_c,F(n_{f_c}))_i$. The indeterminates are given by the error
value vector corresponding to the error pattern.

\item Suppose that the decoder finds a solution to one of these
system of equations. Compute $d_i=P_m+e_{d_i}$; If none of these
systems of equations has a solution, set $d_i=P_m$.

\end{enumerate}

\vspace{-3mm}
\subsection{Condition for errors/failures correction}\vspace{-3mm}
\begin{theorem}\label{theo:hybrid_con}
Suppose there is at most $n_e$ errors and $n_{f_c}$ primary path
failures in the network, both primary path error and protection
path error are possible. The decoding algorithm works at every
node if and only if the column vectors in $A(m_1, m_2)$ are
linearly independent for all 
$A(m_1, m_2) \in \cup_{m \in \{0, \ldots, 2n_e\}}
\mathbf{A}(n_{f_c}+m, 2n_e-m)$.
\end{theorem}
\emph{Proof:} The condition implies that for all $n_c, n_c' \in
\{0, \ldots, n_e\}$ and all possible failure patterns
$F(n_{f_c})$, each member in $\mathbf{A}^F(n_c+n_c',
2n_e-(n_c+n_c'),F(n_{f_c}))$ contains linearly independent
vectors.
The rest of the proof is similar to Theorem \ref{theo:multi_error} and is omitted. \endproof%

The maximum number of vectors contained in each such error pattern
is $4n_e+2n_{f_c}$. Thus, we need at least $M'=4n_e+2n_{f_c}$
equations in \eqref{eq:gen_error_pr} which implies in turn that
$M\geq 4n_e+2n_{f_c}+n_{f_p}$. Since we don't know $n_{f_c},n_{f_p}$ a
priori and in the worse case scenario all failures could happen on the primary paths, we need at least $M=4n_e+2n_f$. On the other hand, using random choice of coefficients
from a large enough field, 
$M = 4n_e+2n_f$ is sufficient to guarantee that the linearly independence condition in Theorem \ref{theo:hybrid_con} satisfies with high probability.

If we restrict the errors/failures to be only on the primary
paths, then the condition becomes each member of
$\mathbf{A}(2n_e+n_f,0)$ is linearly independent and  we can
choose $H$ to be the parity-check matrix of a
$(2n,2n-4n_e-2n_{f})$ RS code. In error/failure value vector $E$,
the locations of the failures are known. The decoding problem can
be viewed as the RS hard decision decoding problem while the
number of error values is bounded by $2n_e$ and the number of
failure values is bounded by $2n_{f}$. It can be done by a
modified BMA \cite{shulin} that works for errors and erasures.

\section{Simulation results and comparisons} \label{sec:result}
In this section, we shall show how our network coding-based
protection scheme can save network resources by some simulations.
Under our adversary error model, when the adversary controls a
single link, one simple protection scheme is to provision three
edge-disjoint paths for each primary connection, analogous to a
(3,1) repetition code. This is referred to as a 2+1 scheme,
meaning that two additional paths are used to protect one
connection. We call our proposed scheme 4+n, i.e., four additional
paths are used to protect $n$ connections. It is expected that
when $n$ becomes large, 4+n will use fewer resources than 2+1. We
provisioned primary and protection paths for both cases and
compared their cost. Our protection scheme can be used in
different networks including optical network deployed in a large
area, or any overlay network no matter what the underlying
supporting network and the scale of the network are.

\begin{figure}[h] \begin{center}
  \subfigure[Labnet03 Network]{\label{fig:Labnet03_topo}\includegraphics[width=120mm, clip=true]{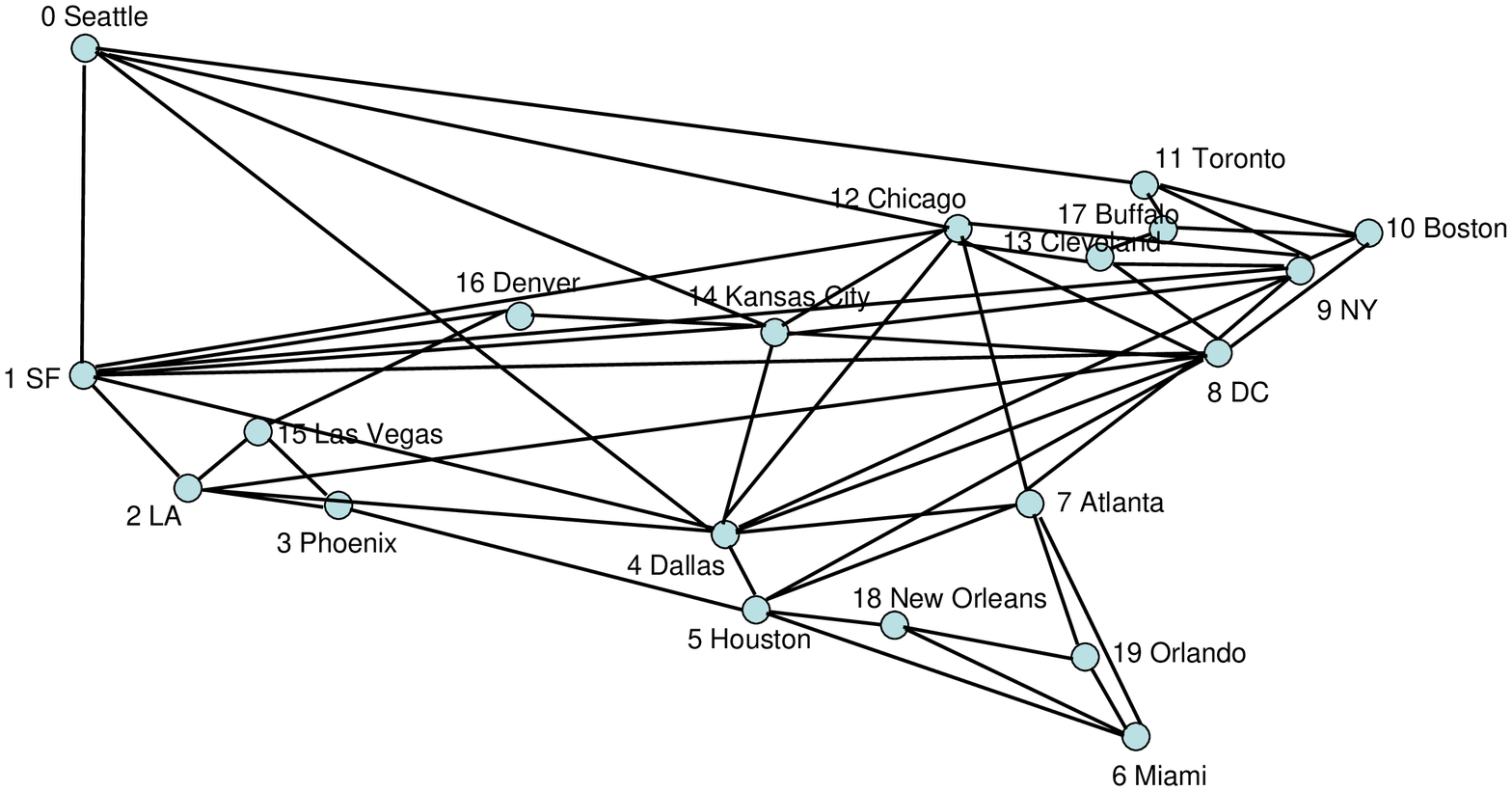}}

  \subfigure[Link costs of Labnet03 network.]{\label{fig:labnet04_wei} \includegraphics[width=100mm, clip=true]{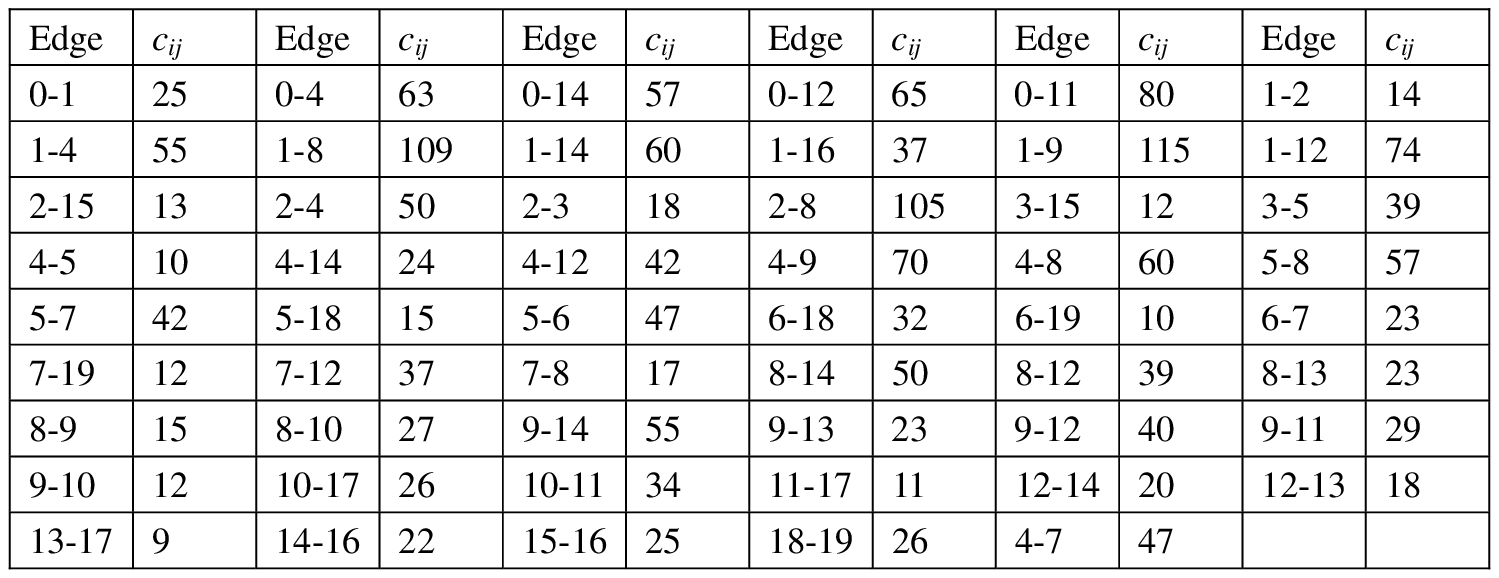}}

 \end{center}
  \caption{\label{fig:Labnet03} Labnet03 network with 20 nodes and 53 edges in North America.}
\end{figure}

\begin{figure}[h]
\begin{center}
\includegraphics[width=80mm, clip=true]{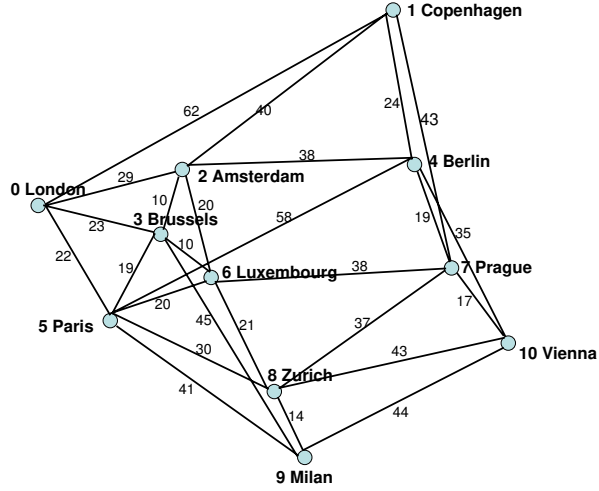}
\caption{\label{fig:COST239} COST239 network with 11 nodes and 26
edges in Europe.}
\end{center}
\end{figure}

\begin{figure}[h]
\centerline{\psfig{figure=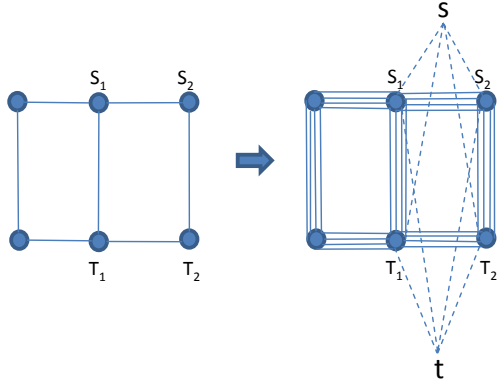,width=3in}}\vspace{-2.5mm}
\caption{\label{fig:infla} Inflation of $G$. The left one is the
original graph $G$. The unicast connections of interest are ${\cal
N} = \{(S_1,T_1),(S_2,T_2)\}$. The right one is the inflated graph
$G'$. }\vspace{-2.5mm}
\end{figure}

In the simulation, we use two networks: 1) Labnet03 network for
North America \cite{menthM05,Walter07} (Fig.\ref{fig:Labnet03}),
2) COST239 Network for Europe \cite{menthM05,kodianG05}
(Fig.\ref{fig:COST239}). Our integer linear programming (ILP)
 for the proposed 4+n scheme is formulated as follows.
The network topology is modelled as an undirected graph $G=(V,E)$.
Considering that usually there are multiple optical fibers between
two cities, we inflate the graph $G$ such that each edge is copied
for several times (four times in our simulations), i.e., there are
four parallel edges between the nodes. An edge $(i,j)$ in $G$ is
replaced by edges $(i,j)_1, (i,j)_2,(i,j)_3,(i,j)_4$ in the
inflated graph. The set of unicast connections to be established
is given in ${\cal N} = \{(S_1,T_1),\ldots,(S_n,T_n)\}$. In order
to model the protection paths as flows, we add a virtual source
$s$ and a virtual sink $t$ to the network and connect $s$ and $t$
with the end nodes of connections in $\cal N$. This procedure is
illustrated in Fig. \ref{fig:infla}. We call this inflated graph
$G'=(V',E')$. Every edge $(i,j)_k$ connecting node $i$ and $j$ is
associated with a positive number $c_{ij}$, the cost of per unit
flow of this link, which is proportional to the distance between
the nodes $i$ and $j$. Assume that each link has enough capacity
so there is no capacity constraint. We hope to find the optimal
$4+n$ paths that satisfy appropriate constraints on the topology
\footnote{we only provision one set of protection paths for
connections in $\cal N$. We could optimally partition $\cal N$
into several subsets, each of which is protected by a set of
protection paths as in \cite{KRLL09}. It will give us better
solution but greatly complicates the ILP. In our simulation, the
4+n scheme shows gains under the simpler formulation. Thus, we
simulate under the simpler formulation. } in the network that
minimize the total cost. One protection path can be viewed as a
unit flow from $s$ to $t$, while one primary path $S_i-T_i$ can be
viewed as a unit flow from $S_i$ to $T_i$. Therefore, the problem
can be formulated as a minimum cost flow problem under certain
conditions. Each edge $(i,j)_k$ is associated with $4+n$ binary
flow variables $f_{ij,k}^m, 1\leq m \leq n+4$, which equals 1 if
path $m$ passes through edge $(i,j)_k$ and 0 otherwise. The ILP is
formulated as follows. \beq \min  \sum_{(i,j)_k\in E'} \sum_{1\leq
m\leq n+4}c_{ij}f_{ij,k}^m . \eeq The constraints are such that
\begin{enumerate}
\item Flow conservation constraints hold for primary paths and
protection paths. \item Each protection path should pass through
the end nodes of all the connections. \item The primary paths are
edge-disjoint. \item The primary paths and the protection paths
are edge-disjoint. \item The protection paths are edge-disjoint.
\end{enumerate}

The minimization is over $f_{ij,k}^{m}, (i,j)_k\in E', 1\leq m
\leq 4+n$ and some auxiliary variables that are used to
mathematically describe the constraints. We assume that when an
adversary attacks an edge in the network she can control all paths
going through that link. Thus, we have edge-disjoint constraints
so that she only causes one path in error in the network. For
detailed mathematical description of the constraints, please refer
to \cite{KRLL09} to see a similar formulation. We call this
formulation as \textbf{ILP1}.

We also provision the paths for 2+1 scheme. The provisioning of
the paths also minimizes the total cost, i.e., the objective is to
minimize $\sum_{(i,j)_k\in E'} (\sum_{1\leq m \leq n} \sum_{1\leq
l \leq 3} c_{ij} f_{ij,k}^{ml})$, where $f_{ij,k}^{ml}$ is the
flow variable for the $l^{th}$ path of the $m^{th}$ primary
connection. Furthermore, the three paths for one connection should
be edge-disjoint. We call this formulation as \textbf{ILP2}.

\begin{figure}[h]
\begin{center}
\includegraphics[width=80mm, clip=true]{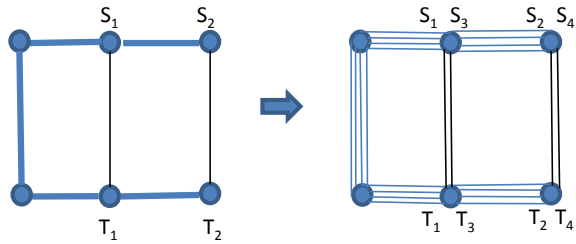}
\caption{\label{fig:UBsolu} A feasible solution of \textbf{ILP1}
is obtained from the optimal solution of \textbf{ILP3}. Here,
${\cal N}_{\frac{1}{2}} = \{(S_1,T_1),(S_2,T_2)\}$ and ${\cal N} =
\{(S_1,T_1),(S_2,T_2),(S_3,T_3), (S_4,T_4)\}$, where $S_1 = S_3,
T_1 = T_3, S_2 = S_4, T_2 = T_4$. Suppose the left graph is the
optimal solution obtained from \textbf{ILP3} on $G$ for ${\cal
N}_{\frac{1}{2}}$. The bold edges indicate that four protection
paths pass through those edges. The right graph is a feasible
solution of \textbf{ILP1} on $G'$. The protection paths are split
into four copies of edges so that the fifth constraint
(edge-disjointness of protection paths) hold. And the paths
$S_1-T_1, S_2-T_2$ are copied to establish $S_3-T_3,S_4-T_4$. It
remains feasible because in $G'$ there are four such paths for
each connection and now we only occupy two of them.}
\end{center}
\end{figure}

However, in general $G'$ contains a large number of edges which
result in a long computation time for \textbf{ILP1}. In order to
simulate and compare efficiently, instead of solving the
\textbf{ILP1} directly, we present an upper bound of the cost for
our proposed 4+n scheme that can be computed much faster. The
connection set $\cal N$ is chosen as follows. Instead of choosing
$n$ connections at random, we choose $n/2$ connections at random
(denoted as the connection set ${\cal N}_{\frac{1}{2}}$) and
duplicate those connections to obtain $\cal N$. So there are two
independent unicast connections between two cities. We remove the
fifth constraint (edge-disjointness of protection paths) from
\textbf{ILP1} and run the ILP instead on the original graph $G$
for ${\cal N}_{\frac{1}{2}}$. We call this ILP as \textbf{ILP3}.
Then, we modify the optimal solution of \textbf{ILP3} properly to
obtain a feasible solution of \textbf{ILP1} for $n$ connections on
$G'$. This is illustrated in Fig. \ref{fig:UBsolu}. The cost of
this feasible solution is an upper bound of the optimal cost of
\textbf{ILP1}. And from the simulation for a small number of
connections we observe that the bound is approximately 10\% larger
than the actual optimal cost. It turns out that solving
\textbf{ILP2} is fast, therefore we obtain the actual optimal cost
for the 2+1 scheme.

In the simulation, we choose $|{\cal N}_{\frac{1}{2}}|$ from 5 to
9 such that $n$ goes from 10 to 18. The ILPs are solved by CPLEX.
The costs for the 4+n scheme and 2+1 scheme are averaged over five
realizations of ${\cal N}_{\frac{1}{2}}$. The average costs and
percentage gains for different number of connections are presented
in Table.\ref{tab:Lab_stat}. and Table.\ref{tab:COST_stat}. As we
expected, the gain of our proposed scheme increases with the
number of connections.

\begin{table}[h]
\begin{center}
\caption{\label{tab:Lab_stat} Comparison of the average costs for
Labnet03 network}
\begin{tabular}{|c|c|c|c|}
\hline $n$ & Average cost for 4+n (upper bound)& Average cost for
2+1 &
Percentage gain \\ \hline 10 &  1826 & 1916.4 & 4.72\% \\
\hline 12 &  2106.4 &  2295.6 & 8.24\% \\ \hline 14 &  2339.6 &
2598.8 & 9.97\% \\ \hline 16 &  2677.6 &  3049.2 & 12.19\% \\
\hline 18 & 3105.2
&  3660 & 15.16\%  \\
\hline

\end{tabular}
\end{center}
\end{table}

\begin{table}[h]
\begin{center}
\caption{\label{tab:COST_stat} Comparison of the average costs for
COST239 network}
\begin{tabular}{|c|c|c|c|}
\hline $n$ & Average cost for 4+n (upper bound) & Average cost for
2+1 &
Percentage gain \\ \hline 10 &  1226 &  1245 & 1.53\% \\
\hline 12 &  1548 &  1628.4 & 4.94\% \\ \hline 14 & 1742.4 &  1854
& 6.02\% \\ \hline 16 &  1810.8 &  1958.4 & 7.54\% \\ \hline 18 &
1883.2 & 2114.4 & 10.93\% \\ \hline

\end{tabular}
\end{center}
\end{table}

Intuitively, our proposed scheme will have more gain when the
connections are over long distances, e.g., connections between the
east coast and the west coast of the US. Roughly speaking, the
number of paths crossing the long distance (inducing high cost) is
$4+n$ for our scheme, while it is $3n$ for the 2+1 scheme. We also
ran some simulation on Labnet03 network to verify this by choosing
the connections to cross the America continent. For a ten
connections setting, we observed 36.7\% gain. And when $n=6$ and
$n=7$, we observed up to 15.5\% and 17.8\% gains respectively. We
conclude that our 4+n scheme is particularly efficient in
allocating network resources when the primary paths are over long
distances or have high cost.

\section{Conclusions and Future Work}\label{sec:conclusion}
In this paper we considered network coding based protection
strategies against adversarial errors for multiple unicast
connections that are protected by shared protection paths. Each
unicast connection is established over a primary path and the
protection paths pass through the end nodes of all connections. We
demonstrated suitable encoding coefficient assignments and
decoding algorithms that work in the presence of errors and
failures. We showed that when the adversary is introducing $n_e$
errors, which may be on primary paths or protection paths, $4n_e$
protections are sufficient for data recovery at all the end nodes.
More generally, when there are $n_e$ errors and $n_f$ failures on
primary or protection paths, $4n_e+2n_f$ protection paths are
sufficient for correct decoding at all the end nodes. Simulations
show that our proposed scheme saves network resources compared to
the 2+1 protection scheme, especially when the number of primary
paths is large or the costs for establishing primary paths are
high, e.g., long distance primary connections.

Future work includes investigating more general topologies for
network coding-based protection. The 2+1 scheme can be viewed as
one where there is usually no sharing of protection resources
between different primary connections, whereas the 4+n scheme
enforces full sharing of the protection resources. Schemes that
exhibit a tradeoff between these two are worth investigating. For example, one could consider provisioning two primary paths for each connection, instead of one and design corresponding network coding protocols. This would reduce the number of protection paths one needs to provision, and depending on the network topology, potentially have a lower cost. It
is also interesting to further examine the resource savings when
we partition the primary paths into subsets and provision
protection resources for each subset separately. Furthermore, in
this paper we considered an adversarial error model. When errors
are random, we could use classical error control codes to provide
protection. But it is interesting to consider schemes that combine
channel coding across time and the coding across the protection
paths in a better manner. A reviewer has pointed out that rank
metric codes \cite{SilvaKK08J} might be also useful for this
problem.

\bibliographystyle{IEEEtran}
\bibliography{Protect}

\end{document}